\documentclass[12pt]{JHEP3}

\usepackage[centertags]{amsmath}
\usepackage{amssymb}

\newcommand{\beq}{\begin{equation}}
\newcommand{\eeq}{\end{equation}}
\newcommand{\beqa}{\begin{eqnarray}}
\newcommand{\eeqa}{\end{eqnarray}}

\newcommand{\half}{\frac{1}{2}}

\newcommand{\Tr}{{\text{Tr}}}

\title{Superconformal Indices for ${\cal N}=6$ Chern Simons Theories}

\author{Jyotirmoy Bhattacharya and Shiraz Minwalla\\
\small{\emph{Department of Theoretical Physics,
                   Tata Institute of Fundamental Research,}}\\
\small{\emph{Homi Bhabha Rd, Mumbai 400005, India.}}  \\
E-mail:\ \ {\bf jyotirmoy@theory.tifr.res.in, minwalla@theory.tifr.res.in}

}

\abstract{
Aharony, Bergman, Jafferis and 
Maldacena have recently proposed a dual gravitational description for a 
family of superconformal Chern Simons theories in three spacetime dimensions. 
In this note we perform the one loop computation that determines the field 
theory superconformal index of this theory and compare with the index computed over the 
Fock space of dual supersymmetric gravitons. In the appropriate limit 
(large $N$ and large $k$) we find a perfect match.
}

\preprint{
  TIFR/TH/08-53 \\
    \texttt{arXiv:0806.3251 [hep-th]}
}

\begin{document}

\maketitle

\section{Introduction}

Aharony, Bergman, Jaffiers and Maldacena (ABJM) have recently proposed that a 
class of $d=3$ $U(N)\times U(N)$  ${\cal N}=6$ superconformal Chern Simons 
field theories admit a dual description in terms of M theory 
compactified on $AdS_4 \times {S^7 \over Z_k}$ \cite{Aharony:2008ug}. 
This correspondence has been further studied in \cite{Benna:2008zy}.
The theories studied by these authors are parameterized by two integers $N$ and $k$. 
In dual bulk terms $k$ is the rank of the orbifold action on $S^7$ while $N$ represents 
the number of units of 7 form flux that pierce $S^7/Z_k$. In field theory 
terms $N$ denotes the rank of each of the $U(N)$ factors of the gauge group 
and $\pm k$ are the levels of the Chern Simons terms associated with 
each of these gauge groups.

With an appropriate normalization for fields, the effective 't Hooft 
coupling constant of any large $N$ gauge theory is given by $N$ times the 
inverse of the coefficient of the action. As the coefficient of the 
Chern Simons term is proportional to $k$, the effective 't Hooft coupling 
of the ABJM field theory is proportional to $N/k$. As $N$ and $k$ are both 
integers this 't Hooft coupling cannot be varied continuously; indeed a 
shift in $k$ by unity shifts $\lambda $ by the discrete amount $\delta \lambda =-{\lambda^2 \over N}$.  
Note however  that $\delta \lambda \to 0$ in
the 't Hooft limit ( $N \to \infty$ with 
$\lambda$ held fixed). Consequently $\lambda$ is effectively 
a continuous parameter in the 't Hooft limit. It follows that the superconformal Witten 
index, defined for arbitrary 3 dimensional superconformal field theories 
in \cite{Bhattacharya:2008zy},
must be invariant under deformations of $\lambda$ in this 't' Hooft 
scaling regime\footnote{
This is only true of contributions to the index from states 
whose energy stays finite as $N$ is taken to infinity.}. 

In this note we compute the superconformal index (as defined in \cite{Bhattacharya:2008zy}) of 
the ABJM theory in two different regimes. We first use the techniques of 
\cite{Sundborg:1999ue,Aharony:2003sx,Aharony:2005bq,Kinney:2005ej,Nakayama:2005mf,Gaiotto:2007qi} to 
find 
an expression for this index at $k=\infty$ (and so $\lambda=0$ but at arbitrary
$N$) in terms of an integral over two $N \times N$ unitary matrices. 
Taking a further large $N$ limit we are able to evaluate these unitary 
integrals explicitly using saddle point techniques.

Next we evaluate the index of this theory at infinite $N$ and large 
$\lambda$, using the ABJM proposal for the dual description of this theory. 
Effectively, we compute the index over the Fock space of 
non interacting supersymmetric $U(1)$ neutral (see below) 
gravitons in $AdS_4 \times {S^7\over Z_k}$.
We perform this calculation explicitly in the 't Hooft limit, but also explain the 
generalization of this calculation to  finite values of $k$, and so to values of $\lambda$ that 
scale like $N$ in the large $N$ limit.

We find that our two independent computations of the superconformal Witten 
index of \cite{Bhattacharya:2008zy} agree perfectly in the 't Hooft limit. We view this agreement as 
a test of the ABJM proposal. Indeed, since the index computed in this  
paper is the most general superconformal index \cite{Bhattacharya:2008zy} for a ${\cal N}=6$ 
superconformal field theory \footnote{By a superconformal index we mean a quantity 
whose invariance under marginal deformations is guaranteed by superconformal 
invariance alone.} our calculation verifies the most detailed matching of 
supersymmetric states predicted by ABJM conjecture taken together with 
the requirement of superconformal invariance alone. Of course the agreement 
of the two independent index computations reported in this paper is closely 
related to the agreement of the spectrum of chiral operators of the field 
theory and the spectrum of gravitons in $AdS_4 \times {S^7\over Z_k}$ 
reported in \cite{Aharony:2008ug}. Note, however, that while the chiral 
ring described by ABJM counts states described by at least four supercharges 
(two supersymmetries and their Hermitian conjugates) the index constructed
in this paper receives contributions from all states that are annihilated 
by a minimum of two supercharges. Note also that the invariance under 
$\lambda$ deformations of the index computed in this paper follows from 
superconformal invariance alone, and uses no dynamical information about the 
field theory beyond its field content. In particular our field theoretic 
computation of the index is blind to the nature of the ABJM superpotential 
whose form played an important
role in the construction of the ABJM chiral ring \cite{Aharony:2008ug}.

The matching of the index computed perturbatively in the ABJM theory with the index 
computed over the spectrum of bulk gravitons is reminiscent of a similar match in 
the case of ${\cal N}=4$ Yang Mills theory \cite{Kinney:2005ej}. As explained in that 
paper, $AdS_5\times S^5$ hosts a family of 1/16 BPS black holes which the 4 dimensional 
superconformal index appears to be blind to. In a similar fashion the 3 dimensional superconformal \
index computed in this paper appears to be blind to the 1/12 BPS black holes presumably 
hosted by the $AdS_4 \times \frac{S^7}{Z_k}$ dual background 
(see \cite{Kostelecky:1995ei,Chong:2004na,Cvetic:2005zi}).

This note is organized as follows. In \S \ref{fieldcal} below we review the symmetry 
algebra of the ABJM theory, the definition of the Witten index of \cite{Bhattacharya:2008zy} for 
this theory and the field content of the ABJM Chern Simons theory. We then 
perform a one loop field theory computation to present a field theoretic 
formula for this index in terms of an integral over two unitary matrices 
and evaluate these integrals in the large $N$ limit. In \S \ref{gravcal}
we present our computation of the same index over the spectrum of 
$U(1)$ invariant multi gravitons in $AdS_4 \times {S^7\over Z_k}$. In 
\S \ref{dis} we end with a discussion of our results and the generalizations 
they suggest.

\section{Field Theory computation of the index} \label{fieldcal}

The symmetry algebra of the ABJM theory is the $d=3$ ${\cal N}=6$ 
superconformal algebra. The structure of this algebra and its unitary 
representations has been reviewed in detail in \cite{Bhattacharya:2008zy}, and we will use the 
notation of that paper in what follows. 

The bosonic subgroup of the $d=3$ ${\cal N}=6$ superconformal algebra is 
$SO(3,2) \times SO(6)$. All states and operators in this theory are labeled  
by their quantum numbers under the maximal compact subalgebra of the 
superconformal algebra, $SO(3) \times SO(2) \times SO(6)$. In what follows
we will denote the eigenvalue of the Cartan generator of $SO(3)$ by $j$, 
the eigenvalue under $SO(2)$ (the scaling dimension or global $AdS_4$ energy)
by $\epsilon_0$ and the three Cartan generators of $SO(6)$- defined as the 
eigenvalues under the generators of rotations in the three orthogonal two 
planes - as $h_1,h_2, h_3$. \footnote{$SO(6)$ may be thought of as the group 
of rotations about the origin in $R^6$ parameterized by $x^i$, $i=1 \ldots 6$. 
$h_1$, $h_2$ and $h_3$ are simply the generators of rotations in the two planes 
(12), (34) and (56) respectively. Throughout this paper we will label representations 
of $SO(6)$ by their highest weights under $(h_1, h_2, h_3)$. In our conventions 
an $SO(6)$ weight is positive if $h_1$ is positive, or if $h_1=0$ and $h_2$ is positive 
or if $h_1=h_2=0$ and $h_3$ is positive. We use a similar conventions for $SO(8)$ representations
below.} 

The twelve supercharges of the ABJM theory each have $\epsilon_0=\half$, 
and transform in the $j=\half$ representation of $SO(3)$ algebra and the 
vector $(h_1, h_2, h_3)=(1,0,0)$ of $SO(6)$ algebra. The only propagating fields in the ABJM theory 
are a set of bi-fundamental and anti bi-fundamental scalars and fermions. 
All scalars have dimension
$\epsilon_0 =\half$ and are scalars under $SO(3)$, while all fermions have 
dimension $\epsilon_0=1$ and transform in the spin half representation 
of $SO(3)$. Bi-fundamental scalars/fermions transform in the 
$(\half, \half, \pm \half)$ representation of $SO(6)$, while anti 
bi-fundamental scalars and fermions transform in the 
$(\half, \half, \mp \half)$ representation of $SO(6)$. The symmetry 
transformation properties of the supersymmetries, propagating fields and 
derivatives of the ABJM theory are listed in table 1 below. 
In table 1 and throughout this paper, the symbol $\phi_{12}$ and $\psi_{12}$ 
respectively denote scalar and fermionic fields that transform in the fundamental 
of the first $U(N)$ gauge group  and 
antifundamental of the second $U(N)$ gauge group, while $\phi_{21}$ and $\psi_{21}$ 
respectively denote scalar and fermionic fields which transforms in the antifundamental of the first 
$U(N)$ gauge group and in the fundamental of the second.

\begin{table}\label{ABJMfldcont}
\caption{A list of the field content of ABJM theory, the supercharges, the derivatives and the representations.}
\begin{center}
\begin{tabular}{|c|c|c|c|c|}
\hline
type of  & operators & scaling & $SO(3)$  & $SO(6)$ \\
operators&&dimension ($\epsilon_0$) &highest weight& highest weight \\
\hline
&&&&\\
& $\phi_{12}$ & $\half$ & $0$ & $(\half,\half,-\half)$\\
&&&&\\
dynamical& $\psi_{12}$ & $1$ & $\half$ & $(\half,\half,\half)$\\
fields&&&&\\
& $\phi_{21}$ & $\half$ & $0$ & $(\half,\half,\half)$\\
&&&&\\
& $\psi_{21}$ & $1$ & $\half$ & $(\half,\half,-\half)$\\
&&&&\\

\hline
supersymmetry &&&& \\
generators & Q & $\half$ & $\half$ & $(1,0,0)$ \\
&&&& \\
\hline
&&&& \\
derivatives & $\partial$ & $1$ & $1$ & $(0,0,0)$ \\
&&&& \\
\hline
\hline

\end{tabular}
\end{center}
\end{table}

In this note we will compute the Witten index 
\begin{equation}\label{inddef}
I^{W} = \Tr \left( (-1)^F x^{\epsilon_0+j} y_1^{h_2} y_2^{h_3} \right)
\end{equation}
for the ABJM theory quantized on $S^2 \times R$. As explained in 
\cite{Bhattacharya:2008zy} this index receives contributions 
only from states that are annihilated by a special supercharge $Q$ together
with its Hermitian conjugate $Q^\dagger$. $Q$ has quantum numbers 
$\epsilon_0=\half$, $j=-\half$, $(h_1,h_2,h_3)=(1,0,0)$. 
It follows from the superconformal algebra that 
$$\{Q, Q^\dagger \}= \left( \epsilon_0-j-h_1 \right) \equiv  \Delta.$$ 
As a consequence, a state is annihilated by both $Q$ and $Q^\dagger$ if and 
only if $\Delta =0$. Consequently the index \eqref{inddef} 
receives contributions only from states with $\Delta=0$. In table 2 below 
we list all $\Delta=0$ propagating fields and derivatives of the ABJM theory.
We also list the partition function over all $\Delta=0$  bosonic fields   
and their derivatives ($f_{12}^{\text{bosonic}}$ 
and $f_{21}^{\text{bosonic}}$), the partition function over $\Delta=0$ 
fermionic fields and their derivatives ($f_{12}^{\text{fermionic}}$ 
and $f_{21}^{\text{fermionic}}$), and the Witten index ($f_{12}$ and $f_{21}$ ) over these
fields.

\begin{table}
\caption{List of the supersymmetric ($\Delta = 0$) fields or `letters' of the theory over which we calculate the index}
\begin{center}
\begin{tabular}{|c|c|c|c|c|c|c|}
\hline
letter&$\epsilon_0$  & $SO(3)$  & $SO(6)$ & $\epsilon_0$+j & partition& index \\
&& (j)& weights&&function& \\
\hline
bi-fundamental&&&&&&  \\
\cline{1-1}

&&&&&& \\

$(\phi_{12})_1$& $\half$ & $0$ & $(\half,\half,-\half)$ & $\half$ & $f_{12}^{\text{bosonic}} = $& \\
&&&&&$\frac{x^{\half}}{1-x^2}\left( \frac{y_1}{y_2}+\frac{y_1}{y_2}\right)$& \\
$(\phi_{12})_2$ & $\half$ & $0$ & $(\half,-\half,\half)$ & $\half$ &&$f_{12} = $ \\
&&&&&&$\frac{x^{\half}}{1-x^2}\left( \frac{y_1}{y_2}+\frac{y_1}{y_2}\right)$ \\
\cline{2-6}
&&&&&& $-\frac{x^{\frac{3}{2}}}{1-x^2}\left( y_1 y_2 + \frac{1}{y_1 y_2}\right)$\\

$(\psi_{12})_1$& $1$ & $\half$ &$(\half,\half,\half)$& $\frac{3}{2}$ &$f_{12}^{\text{fermionic}} = $& \\
&&&&&$\frac{x^{\frac{3}{2}}}{1-x^2}\left( y_1 y_2 + \frac{1}{y_1 y_2}\right)$& \\
$(\psi_{12})_2$& $1$ & $\half$ &$(\half,-\half,-\half)$& $\frac{3}{2}$ && \\
&&&&&& \\
\hline
anti&&&&&&  \\
bi-fundamental&&&&&&  \\
\cline{1-1}
&&&&&& \\
$(\phi_{21})_1$& $\half$ & $0$ &$(\half,\half,\half)$& $\half$ &$f_{21}^{\text{bosonic}} = $& \\
&&&&&$\frac{x^{\half}}{1-x^2}\left( y_1 y_2 + \frac{1}{y_1 y_2}\right)$& \\
$(\phi_{21})_2$& $\half$ & $0$ &$(\half,-\half,-\half)$& $\half$ && $f_{21} = $\\
&&&&&&$\frac{x^{\half}}{1-x^2}\left( y_1 y_2 + \frac{1}{y_1 y_2}\right)$ \\
\cline{2-6}
&&&&&&$-\frac{x^{\frac{3}{2}}}{1-x^2}\left( \frac{y_1}{y_2}+\frac{y_1}{y_2}\right)$ \\

$(\psi_{21})_1$& $1$ & $\half$ &$(\half,\half,-\half)$& $\frac{3}{2}$ &$f_{21}^{\text{fermionic}} = $& \\
&&&&&$\frac{x^{\frac{3}{2}}}{1-x^2}\left( \frac{y_1}{y_2}+\frac{y_1}{y_2}\right)$& \\
$(\psi_{21})_2$& $1$ & $\half$ &$(\half,-\half,\half)$& $\frac{3}{2}$ && \\
&&&&&& \\
\hline
derivative&&&&&& \\
\cline{1-1}
&&&&&& \\
 $\partial$ & $1$ & $+1$ & $(0,0,0)$ & $2$ &&\\
&&&&&& \\
\hline
\hline

\end{tabular}
\end{center}
\end{table}

Following \cite{Sundborg:1999ue,Aharony:2003sx} it was demonstrated in 
\cite{Nakayama:2005mf} (see equation(2.7)) that the 
free superconformal index of a Yang-Mills theory with the field content 
listed above is given by,
\begin{equation}\label{IndUint}
I^W = \int dU_1 dU_2 \exp \left( \sum_{n=1}^{\infty} \frac{1}{n} [
           f_{12}(x^n,y_{1}^n,y_{2}^n) \Tr U_{1}^n \Tr U_{2}^{-n}  
               + f_{21}(x^n,y_{1}^n,y_{2}^n) \Tr U_{1}^{-n} \Tr U_{2}^{n} ]\right)
\end{equation}

The unitary integrals described in \eqref{IndUint} may be evaluated in the 
large $N$ limit. Let $\rho_n={Tr U_1^n \over N}$ and 
$\chi_n={Tr U_2^n \over N}$. In the large $N$ limit the various $\rho_n$ may
be treated as independent variables (modulo a positivity constraint - see 
for instance \cite{Aharony:2003sx} -  that 
will turn out to be irrelevant for our considerations below) and 
$$DU_1= \prod_n d \rho_n \exp \left( -N^2 \sum_{n}  \frac{\rho_n \rho_{-n}}{n} 
\right), ~~~
DU_2= \prod_n d \rho_n \exp \left( -N^2 \sum_{n}  \frac{\chi_n \chi_{-n}}{n} 
\right)$$
so that
\begin{equation}\label{Indeival}
\begin{split}
I^W = \int \prod_{n \neq 0}  d\rho_n d\chi_n \exp &\left( N^2 \sum_{n=1}^\infty
 \frac{1}{n} 
\left( -|\rho_n|^2 -|\chi_n|^2 
        + f_{12}(x^n,y_{1}^n,y_{2}^n) \rho_n \chi_{-n} \right. \right. \\
& \left. \left.
        + f_{21}(x^n,y_{1}^n,y_{2}^n) \rho_{-n} \chi_{n} \right) \right).
\end{split}
\end{equation}

The integral in \eqref{Indeival} takes the form 
\begin{equation}\label{IndeivalII}
I^W = \int \prod_{n \neq 0}  d\rho_n d\chi_n \exp \left( -N^2 \sum_{n=1}^{\infty}
 \frac{1}{n} \left( (C^n)_i (M^n)_{ij} (C^n)_j \right) \right),
\end{equation}
where the column $C^n$ and the matrix $(M^n)_{ij}$ are given by
\begin{equation}
C^n=\left(\begin{array}{l}
          \chi_n \\
          \rho_n \\
          \chi_{-n} \\
          \rho_{-n} \\
          \end{array}\right),~~~~
M^n=\half \left(\begin{array}{llll}
          0 & 0 & 1 & -f_{21} \\
          0 & 0 & -f_{12} & 1 \\
          1 & -f_{12} & 0 & 0 \\
          -f_{21} & 1 & 0 & 0 \\
          \end{array}\right).
\end{equation}

It is possible to demonstrate that the real part of the quadratic form  $(C^n)_i (M^n)_{ij} (C^n)j $ in \eqref{IndeivalII} is positive whenever the chemical potentials obey the inequalities
\begin{equation}\label{inequality}
x< \text{min} \{ y_1 y_2,~ \frac{y_1}{y_2}, ~\frac{y_2}{y_1}, ~\frac{1}{y_1y_2} \}
\end{equation}
a condition that it necessary for the index to be well defined in the first place \footnote{
In order that the index be well defined it is necessary that every $\Delta=0$ letter contribute to the 
partition function with a weight less than unity, leading to \eqref{inequality}.} As a consequence 
it appears that the integral \eqref{Indeival} is always dominated by the saddle point 
at $\rho_n=\chi_n=0$ for all $n \geq 0$. It thus appears that, in perfect analogy with the situation 
for ${\cal N}=4$ Yang Mills,  the Witten index \eqref{Indeival} never undergoes 
the phase transition into a black hole like phase.

As the saddle point contribution to the integral \eqref{Indeival} vanishes, 
the first nonzero contribution to this integral is given by the inverse 
square root of the determinant 
\begin{equation}\label{indform}
I^W=\prod_{n=1}^\infty \frac{1}{\sqrt{16\det M^n}}
\end{equation}
where the normalization in \eqref{indform} is fixed by the requirement that $I^W$ tends to unity 
when $x=0$ (at which point only the vacuum contributes to the Witten index). 

The determinant is given by,
\begin{equation}
\begin{split}
\text{det}(M^n) &= \frac{1}{16}(1- f_{12}(x^n,y_{1}^n,y_{2}^n) f_{21}(x^n,y_{1}^n,y_{2}^n))^2 \\
&= \frac{1}{16}\frac{\left(1-\frac{x^n}{y_1^n}\right)^2
\left(1-\frac{x^n}{y_2^n}\right)^2
\left(1-x^n y_1^n\right)^2\left(1-x^n y_2^n\right)^2}
{\left( 1-x^{2n}\right)^4}
\end{split}
\end{equation}

so that,
\begin{equation}\label{Indfield}
I^W = \prod_n \frac{\left( 1-x^{2n}\right)^2}
{\left(1-\frac{x^n}{y_1^n}\right)\left(1-\frac{x^n}{y_2^n}\right)
                                                                \left(1-x^n y_1^n\right)\left(1-x^n y_2^n\right)}.
\end{equation}

As the large $N$ unitary integrals in \eqref{Indeival} never undergo a 
large $N$ Gross-Witten-Wadia transition \cite{Gross:1980he, Wadia:1980cp}, it follows that 
the Witten index $I^W$ receives contributions only from states of finite 
energy (and charge) at finite values of the chemical potential. In particular, 
\eqref{Indfield} is blind to states whose energy is of order $N^a$ where 
$a$ is any positive power. 

In order to get a feel for \eqref{Indfield} it is useful to set $y_1=y_2=1$. 
If we define the Indicial entropy $S_{ind}(E)$ by the formula $I^w(x)=\int dE e^{S_{ind}(E)} x^E$
then it is easy to show that $S_{ind}(E) \approx \sqrt{2} \pi \sqrt{E}$ at high energies. This 
is the growth of states of a two dimensional massless gas; a similar growth in density 
of states was captured by the four dimensional index (see \cite{Kinney:2005ej}). This growth is  slower than 
the $E^{\frac{2}{3}}$ growth demonstrated by the index of the M2 brane and M5 brane world volume 
thoeries \cite{Bhattacharya:2008zy}.

\section{Gravity computation of the index} \label{gravcal}

Gravitons\footnote{In this section we use the word `graviton' for any field on $AdS_4$ obtained upon
compactification from a field in the 11 dimensional gravity multiplet.} 
in $AdS_4 \times {S^7}$ may be organized into 
representations of the $d=3$ ${\cal N}=8$ superconformal algebra.
Working in conventions in which the $M2$ brane world volume scalar, fermion
and supersymmetries respectively transform in the 
$(\half, \half, \half, -\half)$, $(\half, \half, \half, \half)$ and 
$(1,0,0,0)$ representations of $SO(8)$, the highest weight states of the 
representations that occur in this decomposition each have $j=0$, 
$\epsilon_0={n \over 2}$ and $SO(8)$ highest weight charges $(n/2, n/2, n/2, -n/2)$. 
See \cite{Bhattacharya:2008zy}, Table 1 for more details.

Gravitons on $AdS_4\times {S^7\over Z_k}$ are  those 
 graviton states on  $AdS_4 \times {S^7}$ whose charge under the 
generator $2h_4$ is 0 mod $k$. \footnote{ As for $SO(6)$ we think of $SO(8)$ as the group of rotations 
in $R^8$ parameterized by $x^i$, $i=1 \ldots 8$.  $h_1, h_2, h_3, h_4$ are the eigenvalues of the generators of rotations in the (12), (34), (56) and (78) planes respectively. We label representations of $SO(8)$ by their highest
weights under $(h_1, h_2, h_3, h_4)$; our positivity convention for weights is the obvious generalization 
of that for $SO(6)$.  Note that the $Z_k$ orbifolding 
described in  \cite{Aharony:2008ug} is, in our conventions, simply 
a rotation by the angle $4 \pi/k$ in the (78) 2 plane. 
The subgroup of $SO(8)$ that commutes with this rotation is 
$SO(6) \times SO(2)$. The $SO(2)$ factor is simply rotations in the (78) 2 plane itself 
while the $SO(6)$ factor describes rotations among the remaining 
orthogonal 2 planes and is the R symmetry of the surviving ${\cal N}=6$ 
supersymmetry algebra. Note that the supersymmetry of the parent 
theory decomposes into $6_0+1_2+1_{-2}$ under this decomposition, while the 
scalar decomposes into the $4_1+{\bar 4}_{-1}$. } 
In the large $k$ 't Hooft limit under study in this note, all projected in 
gravitons are simply neutral under the $U(1)$ charge $h_4$. Consequently, 
the superconformal index over single gravitons in $AdS_4\times {S^7\over Z_k}$
is simply the projection of the same quantity in  $AdS_4 \times {S^7}$ to 
the sector of zero $h_4$ charge.

The index 
\begin{equation} \label{mtind} 
I^W=\Tr[ (-1)^F x^{\epsilon_0+j}y_1^{h_2} y_2^{h_3} y_3^{h_4}]
\end{equation}
over single gravitions in $AdS_4\times S^7$ was evaluated in 
\cite{Bhattacharya:2008zy} (see equation 2.17 in that paper).
For the convenience of the reader we reproduce the formula here,
\begin{equation}
I^{W}_{AdS_4 \times {S^7}}(x, y_1, y_2, y_3) = 
\frac{\text{Numerator}}{\text{Denominator}},
\end{equation}
where,
\begin{equation} \label{mtind}
\begin{split}
\text{Numerator}
 =& -\sqrt{y_1} \sqrt{y_2} \sqrt{y_3}
   (y_2 y_3
    y_1+y_1+y_2+y_3)
   x^{7/2}\\
   & \quad +\left(y_2 y_3
   y_1^2+\left(y_3 y_2^2+y_3^2
   y_2+y_2+y_3\right)
   y_1+y_2 y_3\right)
   x^3 \\
  & \quad -\left(y_2 y_3
   y_1^2+\left(y_3 y_2^2+y_3^2
   y_2+y_2+y_3\right)
   y_1+y_2 y_3\right)
   x \\
  & \quad +\sqrt{y_1} \sqrt{y_2} \sqrt{y_3}
   (y_2 y_3+y_1
   (y_2+y_3)+1) \sqrt{x} \\
\text{Denominator} =& \left(1-x^2\right)\left(\sqrt{x}
   \sqrt{y_1}
   \sqrt{y_2}-\sqrt{y_3}\right)
   \left(\sqrt{x} \sqrt{y_1}
   \sqrt{y_3}-\sqrt{y_2}\right) \\
    & \quad \left(\sqrt{y_1}-\sqrt{x} \sqrt{y_2}
       \sqrt{y_3}\right) \left(\sqrt{y_1}
       \sqrt{y_2} \sqrt{y_3}-\sqrt{x}\right)
\end{split}
\end{equation}

The index over single gravitons of zero $h_4$ charge is given by 
$$\int \frac{d\theta}{2\pi i} I^{W}_{AdS_4 \times {S^7}}(x, y_1, y_2, 
e^{i \theta}) = 
\int_C \frac{dy_3}{2\pi y_3 i} I^{W}_{AdS_4 \times {S^7}}(x, y_1, y_2, 
y_3)$$
where the contour $C$ surrounds the poles at $y_3=0$, 
$y_3={x y_1  y_2}$ and $\frac{x}{y_1 y_2}$. 
\footnote{The index \eqref{mtind} is well defined only when $x<y_1^a y_2^b y_3^c$ 
where $(a,b,c)$ run over the values of $(2h_1, 2h_2, 2h_3)$ for the antichiral spinor of $SO(6)$.
It follows that the contour of our integral must exclude the poles at $y_3 = \frac{y_1}{x y_2}$ and 
$y_3 = \frac{y_2}{x y_1}$.}

Performing this sum of residues we find that the index over $U(1)$ neutral 
gravitons on $AdS_4\times {S^7\over Z_k}$ is given by,
\begin{equation} \label{indsp}
I^{W}_{\text{Single Particle}}
 = \frac{x }{y_1-x}+\frac{1}{1-x
   y_1}+\frac{x
  }{y_2-x}+\frac{1}{1-x
   y_2}-\frac{2}{1-x^2}.
\end{equation}
a result that is significantly simpler than \eqref{mtind}.

We have also verified \eqref{indsp} more directly. As explained in 
\cite{Pope, Aharony:2008ug} the $U(1)$ neutral gravitons in $AdS_4\times \frac{S^7}{Z_k}$ 
appear in a direct sum representations of the ${\cal N}=6$ superconformal 
algebra labeled by the highest weight states with $\epsilon_0=n$, $j=0$ 
and $(h_1,h_2, h_3)=(n,n,0)$ for $n=1 \ldots \infty$. It is not difficult 
to decompose every such representation of the superconformal algebra into 
irreducible representations of the $d=3$ conformal algebra (using, for 
instance, the techniques described in \cite{Bhattacharya:2008zy}). This decomposition 
could also be read off from the Table 1 in \cite{Pope}.
In Table 3 we list those conformal representations that have states 
with $\Delta=0$. Only states with maximum values of $h_1$ and $j$ in the
representations listed in Table 3 have $\Delta=0$ and contribute to the index. 
Consequently, the contribution of any of the  representations listed below, 
to the index is simply given by 
$$I^W_{\epsilon_0,j,h_2,h_3}=(-1)^j \frac{x^{\epsilon_0 + j}}{1-x^2} 
\chi_{SO(4)}^{(h_2, h_3)}(y_1, y_2)$$ 
where $\epsilon_0$ and $j$ are respectively the dimension and $SO(3)$ charge
of the highest weight state in the representation, the factor of $\frac{1}
{1-x^2}$ is the contribution from the supersymmetric derivatives and 
$\chi_{SO(4)}^{(h_2, h_3)}(y_1, y_2)$ is the $SO(4)$ character with 
highest weights $h_2, h_3$. Summing this quantity over all the representations
listed in Table 3  for all $n\geq 1$ we recover \eqref{indsp}.

\begin{table}\label{gravspec}
\begin{center}
\begin{tabular}{|c|c|c|c|c|c|}
\hline
Range of n & Scaling & $SO(3)$ & $SO(6)$ highest weight& $\epsilon_0 + j$ & statistics\\
&dimension $\epsilon_0$ & weight (j) & (orthogonal basis)& & \\
\hline
$n \geq 1$ & $n$ & $0$ & $(n,n,0)$ & $n$ & bosonic\\
$n \geq 1$ & $n+\half$ & $\half$ & $(n,n,1)$ & $n+1$ & fermionic \\
$n \geq 1$ & $n+\half$ & $\half$ & $(n,n,-1)$ & $n+1$ & fermionic\\
$n \geq 1$ & $n+\half$ & $\half$ & $(n,n-1,0)$ & $n+1$ & fermionic\\
$n \geq 1$ & $n+1$ & $1$ & $(n,n,0)$ & $n+2$ & bosonic\\
$n \geq 2$ & $n+1$ & $1$ & $(n,n-1,1)$ & $n+2$ & bosonic\\
$n \geq 2$ & $n+1$ & $1$ & $(n,n-1,-1)$ & $n+2$ & bosonic\\
$n \geq 2$ & $n+\frac{3}{2}$ & $\frac{3}{2}$ & $(n,n-1,0)$ & $n+3$ & fermionic \\
\hline
\hline
\end{tabular}
\end{center}
\caption{The supersymmetric ($\Delta =0$) graviton spectrum in $AdS_4 \times \frac{S^7}{Z_k}$. Here
n is an integer greater than or equal to 1.}
\end{table}

In order to compute the index over multi gravitons we use the 
single partice index \eqref{indsp} and the formulas of 
Bose statistics to obtain \footnote{According to the rules of Bose Statistics
a single particle index of the form $\sum_n c_n x^n$ translates into a multi
particle index $\prod_n (1-x^n)^{-c_n}$ where $c_n$ are integers that could 
be either negative or positive, and $x$ schematically represents all 
chemical potentials.}
\begin{equation}
I^W = \prod_n \frac{\left( 1-x^{2n}\right)^2}
{\left(1-\frac{x^n}{y_1^n}\right)\left(1-\frac{x^n}{y_2^n}\right)
 \left(1-x^n y_1^n\right)\left(1-x^n y_2^n\right)}.
\end{equation}
in perfect agreement with \eqref{Indfield}.

\section{Discussion} \label{dis}

In this note we have computed the supersymmetric index of 
\cite{Bhattacharya:2008zy} for the ABJM theory in two different ways. 
We first performed a one loop field theory computation to evaluate this 
index in the free theory field at large $N$; this calculation was performed 
at $\lambda=\frac{N}{k}=0$. We then evaluated the same index over the Fock 
space of $U(1)$ neutral gravitons in $AdS_4 \times \frac{S^7}{Z_k}$. This 
calculation was valid in the large $N$ limit with $\lambda$ fixed at a large
value. The results of our two calculations match perfectly, providing a 
check of the ABJM conjecture. 

It would be easy to generalize the gravitational calculation presented 
in this note to apply the large $N$ limit with $k$ held fixed. All one needs
to do is to project \eqref{mtind} onto the sector with $2h_4=0$ mod $k$ 
(rather than simply to zero), before applying the formulas of Bose statistics.
For instance when $k=1$ this simply amounts to setting $y_3$ to unity in  
\eqref{mtind}. It may be possible to reproduce the full finite $k$ 
gravitational index from an (almost) free field theory calculation after 
summing over flux sectors on $S^2$, as suggested by the discussion in  
\cite{Aharony:2008ug}. It would be very interesting to try to carry this through.

Another direction that would be interesting to explore would be the 
determination of the full supersymmetric partition function (rather than 
the supersymmetric index) of the ABJM theory, with varying amounts of 
supersymmetry. For instance a formula for finite $N$ partition function over 
the chiral ring (states that preserve 4 supercharges) has been proposed 
in \cite{Bhattacharyya:2007sa} at $k=1$. It would be interesting to verify this formula and to generalize 
it to other values of $k$. More ambitiously one could attempt to determine the full 
partition function (in contrast to the index computed in this paper) over all supersymmetric 
of in the ABJM theory; note however that this programme has not yet been completed even 
for ${\cal N}=4$ Yang Mills (see \cite{Grant:2008sk} for a recent status report).  

In this connection note that like $AdS_5\times S^5$, the ABJM gravitational 
background presumably hosts supersymmetric black holes 
that preserve 2 supersymmetries (see \cite{Kostelecky:1995ei,Chong:2004na,Cvetic:2005zi} for 
relevant work). This is exactly the minimum amount of 
supersymmetry that a state needs to preserve to contribute to the 
index described in this paper. However the index computed in this paper 
sees no sign of these states. Indeed a naive estimate suggests that the 
entropy of these supersymmetric black holes scale like $N^2/\sqrt{\lambda}$
times functions of chemical potentials. As a result, the entropy of these 
black holes appears to be a function of $\lambda$ at large $\lambda$, and so 
cannot be captured by any quantity like an index that is independent of 
$\lambda$. The smooth dependence of the entropy of a supersymmetric 
configuration on a continuous coupling constant appears non intuitive at 
first sight, and it would be interesting to understand how this comes about. 
Perhaps the states that make up the entropy of the black hole receive 
important contributions from the nontrivial flux sectors (these sectors, 
whose energy scales like $N$, could in principle contribute to the 
entropy of a black hole - whose energy scales like $N^2$, even in the 
't Hooft limit). \footnote{We thank R. Gopakumar for this suggestion.}

Turning to the spectrum of nonsupersymmetric states in this theory, it seems possible that 
the integrability of the spin chain spectrum of chiral operators 
could carry over to the ABJM theory. It would be interesting to
study this possibility in more detail. Indeed, one of the exciting aspects of the 
ABJM proposal (in our opinion) are the prediction that the effective 
string that describes spin chain dynamics metamorphoses into a membrane 
at $\lambda ={\cal O}(N)$. It would be very interesting to attempt to get 
a concrete handle on this. 

Finally, of course the ABJM duality permits the computation of all correlators 
(not just the spectrum) of all the chiral operators in the theory at strong coupling. 
A simple scaling estimate reveals that 
$k$  point functions of these operators scale like $(\lambda^{\frac{1}{4}}/N)^{k-2}$ 
at strong coupling. In particular, three point functions are functions 
of $\lambda$ and cannot enjoy the nonrenormalization properties of 3 point 
functions of chiral operators in ${\cal N}=4$ Yang Mills theory in $d=4$ \cite{Lee:1998bxa}.

\section*{Acknowledgements}
We would especially like to thank S. Bhattacharyya and R. Loganayagam for several useful discussions 
during the course of this project. We would also like to thank M. Berkooz, N. Drukker,
B. Ezhuthachan, M. Gaberdiel, R. Gopakumar, S. Kim, G. Mandal,  S. Mukhi, C. Papageorgakis, S. Raju,
S. Wadia, all the participants of Monsoon Workshop on String Theory and all the students in the TIFR 
theory room for useful discussions and comments on the manuscript.
The work of SM was supported in part by a Swarnajayanti Fellowship.
We would like to acknowledge our debt to the steady and generous support of the people of 
India to research in the basic sciences.

\bibliographystyle{style}
\bibliography{bibfile}

\end{document}